# Admissions Criteria and Diversity in Graduate School

Casey W. Miller
Associate Professor and Associate Director of Graduate Programs
Physics Department, University of South Florida,
4202 East Fowler Ave, Tampa, FL 33620

Abstracto. In this work, I point out the negative implications for diversity in graduate school resulting from the use of cutoff scores on the GRE in the admissions process. In light of the data presented, as well as a swelling body of evidence suggesting no long term correlation with research success, I pose several challenges to the community related to the continued use of the GRE.

The development of the APS Bridge Program (APS-BP) [1] provides our community with a great opportunity to help physics shed the notoriety of being the least diverse of the sciences. We can all play a role in this effort, whether or not our individual programs are or will become affiliated with the APS-BP. In this article, I hope to raise awareness of some relatively simple but impactful means to enhance racial and gender diversity. What follows is hardly comprehensive, but hopefully suggests pragmatic next steps that can be widely and rapidly implemented.

There are about 180 physics programs listed in the AIP Graduate Programs book. The General GRE is required by 96%; a quarter of these have an explicitly stated minimum Quantitative GRE score for admission, with the median stated cut-off being 700 ($64^{th}$ – $70^{th}$ percentile, depending on year). As educators, we naturally expect exams to be meaningful. Most people *believe* this is the case for the GRE exams, and may thus prefer high scores. But analysis of the data often finds no significant correlation between long-term success and GRE scores (the stated predictive power is limited to first year grades). In my time on the Graduate Council of the University of South Florida, I have seen few programs that have taken the time to correlate "success" with admissions materials. Of these, none, including my program, have concluded the GRE predicts success in research, which no one argues is the aim of the PhD. Our analysis finds that the QGRE correlates with only one metric, the graduate GPA (but it is such a weak correlation the scientist in me rebels when fitting it to a line). That said, we find undergraduate GPA to be a better predictor of graduate GPA. We also find that undergraduate GPA is correlated with all three sections of the General GRE.

So why use the GRE at all? One certain answer: national rankings. Consider *US News*, whose rankings of graduate programs are widely influential among both prospective graduate students and administrators. In the *US News* formula, the weight given to the mean GRE score is 12% (6% each to Verbal and Quantitative). The acceptance rate weighting of 6% is added to give "student selectivity" a total weight of 18%. This exceeds student/faculty ratio (4.5%), percent of faculty with awards (2.5%), and doctoral degrees granted (5%), and is far too close in my opinion to tangible productivity measures such as research expenditures (15% each for dollars per program and dollars per faculty member). This weighting can lead administrators to the simple conclusion that their rankings will go up if they only admit students with perfect GRE scores.

Justifying using the GRE becomes significantly more complicated, however, when the test results are dissected by race and gender. Figure 1 plots QGRE scores by race/ethnicity and gender for US citizens whose intended graduate major was "physical sciences". The top and bottom of the lines are the $75^{th}$ and $25^{th}$ percentiles of the score distributions, respectively; the tick is the mean; the number of test takers is noted near the data labels. This pattern is qualitatively unchanged when controlling for undergraduate GPA. Note the implications for diversity of using 700 as a minimum acceptable score: nearly three quarters of Hispanics would be rejected, and significantly more than this for American Indians, African Americans, and Puerto Ricans; similarly, women are filtered out at a higher rate than men. Mixing cut-off scores with these racial and gender disparities sets the foundation of a glass ceiling erected by the lopsided treatment of minorities and women before they even step foot in grad school.

The Asian>White>Hispanic>Black pattern permeates standardized testing: it is the same for the SAT, and is reflected in the recent race-based levels set by Florida and Virginia for grade schoolers' performance on state-wide standardized tests [2]. The GRE pattern, however, must be more complex in origin than a property tax distribution problem-these are students who have done so well in college that they consider themselves serious enough candidates for graduate studies to spend hundreds of dollars on the tests and score reporting. As for gender, there is not a single section of the GRE general test in which women outscore men, regardless of how the data are diced. Isn't the ubiquity of these patterns odd?

The results of Fig. 1 are not new, but are not trumpeted loudly or often enough to be well known among physicists. When a skeptical colleague asserts, "something is wrong with those data!" I point out that they come from ETS, the company that administers the GRE (nevertheless, even though I accept their validity, I wholeheartedly support the assertion that something is wrong). Others have noted the impact of the GRE on diversity in physics, hopefully not limited to [3-6]; and perhaps of interest to administrators [7].

Do we physicists appreciate how significantly admissions practices can impact diversity? It should be clear that using "minimum acceptable" scores to filter applicants will have adverse unintended consequences. Applicants send many items, and we all aim to take the whole package into consideration. However, I grow curious when programs report average GRE scores that are nearly perfect, particularly given that the published standard measurement error is on the order of 60 points (i.e., 740/$80^{th}$ percentile is insignificantly different from a perfect 800). For the record: the *Guide to the Use of Scores* published by ETS [8] states, "A cut-off score (i.e., a minimum score) should never be used as the only criterion for denial of admission or awarding of a fellowship." The latter sentiment refers, likely, to *Sharif v. New York State Education Department*, in which a federal court found that the sole use of SAT scores for awarding scholarships violated Title IX because of the exam's gender gap. While one hopes cut-off scores are not being intentionally employed for admissions, it is awfully easy to sort a spreadsheet by GRE score and start from the top of the list, giving this metric undue weight. How different is sorting from a cut-off? Can this approach actually provide equal opportunity for admission?

While I'm sure no one fully understands the origins of Fig. 1's data, there are well documented points faculty should note. With other factors equal (e.g., GPA), women score lower on high-pressure timed tests than men. As a result, for example, it is recommended in *Graduate Education in Physics: Which Way Forward?* [9] that time constraints on qualifying exams be eliminated. Minority students often graduate with relatively limited advanced physics coursework. This cannot explain the QGRE results, but is worth noting for programs that use the Physics GRE (for which no equivalent to Fig. 1 is publicly available, to my knowledge). Language must play a role to some extent (probablemente yo no haría un buen examen si estuviera escrito en perfecto español). Finally, there is a strategy for taking the tests that can be learned (would businesses like *The Princeton Review* exist otherwise?) Are any of these points relevant for predicting success in your graduate program?

It is my hope that the biases noted here, along with the emerging body of evidence questioning the utility of the GRE, will nucleate a transformation in admissions practices. To be clear: this is not to suggest that we admit less qualified students for the sake of diversity, but rather a call for us to acknowledge that the historical importance given to GRE scores exceeds its predictive capabilities, and has societal implications that we may not have anticipated. One approach we are pursuing to address this issue involves developing a coarse-grained admissions rubric that ranks applicants based on a variety of factors we have determined to be useful in predicting research success: GPA, recommendations, personal statement, and undergraduate institution (liberal arts students are very successful in our program…please apply!). While we still have more to learn, instituting meaningful admissions practices can help lift physics out of the diversity basement.

In closing, Physics has the opportunity to move forward on the major issue that is diversity in physics. If many parties participate, even with small steps, progress can be accelerated. While none of these will be sufficient to completely address diversity, I would like to propose several challenges to the community:

To PIs: get involved-helping to reform your graduate program is a long-term investment, but your gains will start immediately, e.g., by enabling meaningful broader impacts statements.

To programs: determine (and share!) admissions factors that actually predict research success (the aim of the PhD) and develop corresponding admissions practices; reduce the unintentional weight GRE scores presently receive.

To administrators: encourage and reward appropriate admissions practices (*practices*, not policies).

To program ranking entities: abandon using the GRE in program evaluations-this puts downward pressure on diversity.

To professional societies (AAAS, APS, NSBP, NSHP, SACNAS, AWIS, etc.): develop statements regarding the GRE exams and their usage by graduate programs-your guidance is crucial.

Finally, to the next generation: your voices are more powerful than you think, and are needed more frequently than every four years.

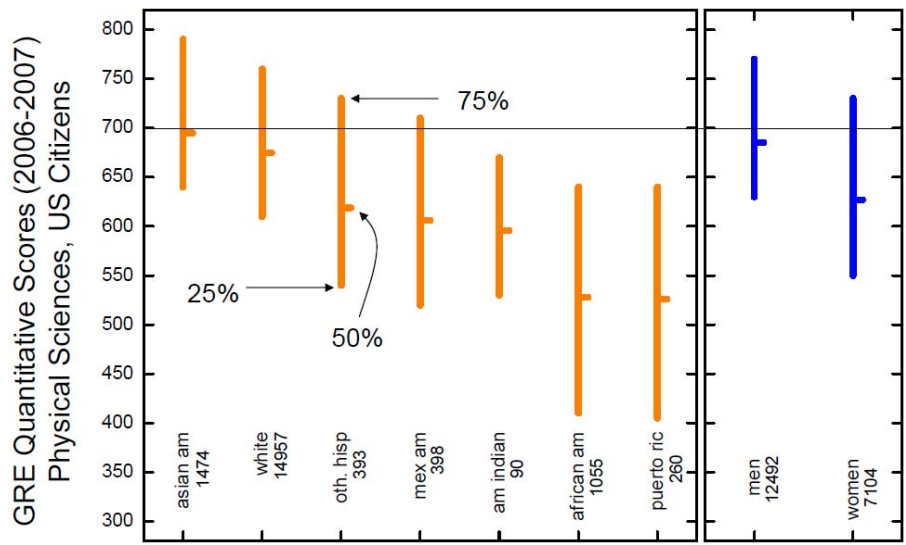

Figure 1. GRE Quantitative scores for test takers self-identifying their intended graduate major as physical sciences (presumably this means astronomy, physics, chemistry, geology, etc.). The bars are intended to show the range of scores (with $25^{th}$, $50^{th}$, and $75^{th}$ percentiles as indicated) for each group, with the number of test takers in each group noted near the labels. The thin horizontal solid black line indicates a score of 700, which is the median stated cutoff score by physics PhD programs that state an explicit cutoff score.